\documentclass[a4paper]{amsart}
\newcommand\stypath{.}
\newcommand\imgpath{img}

\usepackage{\stypath /common}

\usepackage{\stypath /rev}

\usepackage[T1]{fontenc}

\usepackage[russian,english]{babel}

\usepackage[backend=bibtex, style=alphabetic]{biblatex}
\usepackage{chngcntr}
\usepackage{apptools}
\AtAppendix{\counterwithin{thm}{section}}

\usepackage{\stypath /qc}

\usepackage[]{geometry}

\usepackage[foot]{amsaddr}

\colorlet{DRA}{dra}

\title[Alternative QPE using Projection-based Tensor Decompositions]{An \dra{Alternative} Formulation of the Quantum Phase Estimation Using Projection-Based Tensor Decompositions}

\author[Marian Stengl]{\dra{Marian} Stengl${}^{1,2}$}

\address{${}^1$Institut f\"ur Mathematik\\
Technische Universit\"at Berlin\\
Germany}

\email{stengl@math.tu-berlin.de}

\address{${}^2$Zuse Institute\\
Berlin\\
Germany}

\subjclass[2020]{68Q12, 15A69, 47A80}

\keywords{Quantum computing, quantum phase estimation, tensor decomposition, projection operator, circulant matrices}

\date{\today}

\begin{document}
%
%
\begin{abstract}%
\dra{In this paper an} alternative version of the quantum phase estimation \dra{is} \dra{proposed}, in which the Hadamard gates at the beginning are substituted by a quantum Fourier transform.
\dra{This} new circuit coincides with the original one, when \dra{the ancilla is} initialized with $\ket{0}$.
\dra{With} the help of a projection-based tensor decomposition and \dra{closed-form} expressions of its exponential, this new method can be interpreted as \dra{a} multiplier coupled to the Hamiltonian of the corresponding target unitary operator.
Based on this observation a recursive decomposition is \dra{derived}.
\end{abstract}%
%
%
\maketitle
%
%
\section{Introduction}
Quantum Computing has originated in the early to mid 80's in the works by Feynman in \cite{bib:Feynman} as a novel computational scheme using quantum mechanical principles.
Similar ideas have as well been formulated earlier, e.g., in the introduction of \cite{bib:Manin_ComputableNoncomputable}, in \cite{bib:Benioff_MiscroscopicQuantumMechanical} and \cite{bib:Peres_ReversibleLogicQuantumComputer}.
For an introduction to quantum computing the reader is referred to \cite{bib:NielsenChuang}.	

Since then a considerable number of algorithms has been proposed in the past four decades for a wide range of applications such as Grover's algorithms for search problems, (see \cite{bib:Grover}), \dra{integer} factoring \dra{(see \cite{bib:Shor})} \dra{or the solution of systems of linear equations (see \cite{bib:HHL})}.
One of their vital building blocks is the \emph{quantum phase estimation} \dra{(QPE)}.

The original algorithm was \dra{presented} in \cite{bib:Kitaev_QPE}.
We refer here to QPE by the \lq{}textbook\rq{} method as described, e.g., in \cite[Section 5.2]{bib:NielsenChuang}.
Let a unitary operator together with one of its eigenvectors be given.
The aim is the calculation of the phase of the associated eigenvalue, \dra{which is represented by} an ancilla quantum register.

This \dra{article} has two major goals:
The primary \dra{one} is the derivation of an alternative quantum phase estimation \dra{that has more convenient mathematical properties and contains the textbook version as a special case.}
Our method can be interpreted as the operator exponential of a multiplier, which only depends on the number of ancilla qubits used to represent the phase, and the Hamiltonian associated to the aforementioned target operator.
The used techniques are centered around calculus rules for a \dra{class} of projection-based \dra{tensor} decompositions of the Hamiltonian.
\dra{The latter results and their application are the secondary goal of this paper.}

%
%
The rest of this article is organized as follows.
In \cref{sec:np} we introduce the \dra{notation} and gather central results that are used throughout the text.
In \cref{sec:proj} we derive calculus rules for a class of operators that admit a projection-based \dra{tensor} decomposition and give some instructive examples.
\dra{Our} alternative quantum phase estimation is introduced in \cref{sec:alternative_qpe} and the results of the previous section are used to derive a Hamiltonian-based representation.
Additionally, we propose a recursive approach in \cref{sec:recursive_qpe} to decompose our alternative quantum phase estimation.
\section{Notation and Preliminaries}\label{sec:np}
%
%
Let $\cH$ be a finite dimensional, complex Hilbert space.
For \dra{a lowercase} index $n$ we denote $N = 2^n$ and the Hilbert space $\cH_N \simeq \bC^N$ with basis $(\ket{j})_{j = 0}^{N - 1}$.
The latter is the computational base of an $n$-qubit quantum computer.
Clearly we have for $n_0,n_1 \in \bN$ with $n = n_0 + n_1$ and $N_j = 2^{n_j}$, $j = 0, 1$ that $N = N_0 \cdot N_1$ and $\cH_N \simeq \cH_{N_0} \otimes \cH_{N_1}$.

%
%
For a Hilbert space $\cH$ as above we denote the set of all continuous, linear operators on $\cH$ by $\cL(\cH)$.
The identity on $\cH$ is denoted by \dra{$\id_{\cH}$}.
For \dra{$\cH = \cH_N$} we may \dra{just} write \dra{$\id_N = \id_{\cH_N}$}.
Two operators $A, B \in \cL(\cH)$ are said to \emph{commute} if $AB = BA$ is true.

\dra{We} identify operators with their matrix representation in the computational basis unless stated otherwise.
\dra{The indices of vector and matrices start at \emph{zero}.}
Let $A \in \cL(\cH_{N_0})$ and $B \in \cL(\cH_{N_1})$ be given.
Their \emph{Kronecker product} (see \cite[eq. (2.50)]{bib:NielsenChuang}) $A \otimes B \in \cL(\cH_{N_0} \otimes \cH_{N_1})$ is defined by the matrix
\begin{eq*}
A \otimes B =
\left[
\begin{array}{ccc}
a_{0,0} \cdot B & \cdots & a_{0,N - 1} \cdot B\\
\vdots          & \ddots & \vdots\\
a_{N-1,0} \cdot B & \cdots & a_{N-1,N-1} \cdot B
\end{array}
\right].
\end{eq*}
It is also the matrix representation of the tensor product of the two operators.

%
%
Let $N \in \bN$ \dra{and define $\omega := \exp(\ic \frac{2\pi}{N})$}.
\dra{The} \emph{quantum Fourier transform} \dra{is} defined as the matrix $\QFT_N \in \cL(\cH_N)$ with entries $(\QFT_N)_{j,k} = \dfrac{1}{\sqrt{N}}\omega^{j k}$ for $j, k = 0, \dots, N - 1$.
Hence, we have 
\begin{eq*}
\QFT_N
= \frac{1}{\sqrt{N}}
\left[
\begin{array}{ccccc}
1      & 1              & 1                  & \cdots & 1                      \\
1      & \omega         & \omega^2           & \cdots & \omega^{N - 1}         \\
1      & \omega^2       & \omega^4           & \cdots & \omega^{2(N - 1)}      \\
\vdots & \vdots         & \vdots             &        & \vdots                 \\
1      & \omega^{N - 1} & \omega^{2 (N - 1)} & \cdots & \omega^{(N - 1)(N - 1)}
\end{array}
\right].
\end{eq*}
%
%
Let throughout the entire article \dra{be} $t \in \bR$.
For an operator $A \in \cL(\cH)$, we define its \dra{\emph{operator exponential}} (or \dra{\emph{matrix exponential}}), cf. \cite[eq. (2.1)]{bib:Hall_LieGroupsLieAlgebras} by
\begin{eq}\label{eq:defn_exp}
\dra{\exp(A) := \sum_{k = 0}^\infty \frac{1}{k!}A^k \in \cL(\cH).}
\end{eq}
Some natural choices are $t = \pm 1$ and $t = \pm \pi$.
For a unitary operator $U \in \cL(\cH)$ we denote by $\Ham(U) \in \cL(\cH)$ a symmetric, linear, continuous operator such that 
\begin{eq*}
\exp(- \ic t \Ham(U)) = U.
\end{eq*}
This operator is in general not unique.
%
%
The following calculus rules for the operator exponential are of relevance for us.
\begin{lem}[Calculus Rules for the Operator Exponential, cf. \mbox{\cite[Proposition 2.3]{bib:Hall_LieGroupsLieAlgebras}} and \mbox{\cite[Lemma 4.169]{bib:Hackbusch_TensorSpacesNumericalTensorCalculus}}]\label{lem:exp_properties}
The following statements are valid:
\begin{enum}
\item
Let a finite dimensional, complex Hilbert space $\cH$ and commuting matrices $A, B \in \cL(\cH)$ be given.
Then we have
\begin{eq*}
\exp(A + B) = \exp(A)\exp(B).
\end{eq*}
In particular, the operators $\exp(A)$ and $\exp(B)$ commute.
\item 
For all $n \in \dra{\bN}$ and \dra{$A \in \cL(\cH)$} holds
\begin{eq*}
\exp(n A) = \exp(A)^n.
\end{eq*}
\item
For all $A \in \cL(\cH)$ \dra{hold} $\exp(-A) = \exp(A)^{-1}$ and $\exp(A^*) = \exp(A)^*$.
\item
For all $A \in \cL(\cH)$ and all invertible $T \in \cL(\cH)$ with inverse $T^{-1} \in \cL(\cH)$ we have
\begin{eq*}
\exp\left(T^{-1} A T\right) = T^{-1} \exp(A) T.
\end{eq*}
\item
Let two finite dimensional, complex Hilbert spaces $\cH, \cH'$ and linear operators $A \in \cL(\cH)$, \dra{$A' \in \cL(\cH')$} be given.
Then
\begin{eq*}
\exp(A \otimes \id_{\cH'} + \id_{\cH} \otimes A') = \exp(A) \otimes \exp(A') .
\end{eq*}
\end{enum}
\end{lem}

Additionally, we require some results from matrix theory throughout the text, which are introduced here.
A matrix $A \in \cL(\cH_N)$ is called \emph{circulant}, if there exists a sequence $(c_\ell)_{\ell = 0}^{N - 1} \subseteq \bC$ such that for the entries holds $A_{j,k} = c_{\revb{(k - j \pmod{N})}}$ for $j, k = 0, \dots, N - 1$.
Hence, circulant matrices have the form
\begin{eq*}
A = \left[
\begin{array}{ccccc}
c_0       & c_1       & c_2    & \cdots & c_{N - 1}\\
c_{N - 1} & c_0       & c_1    & \cdots & c_{N - 2}\\
c_{N - 2} & c_{N - 1} & c_0    & \cdots & c_{N - 3}\\
\vdots    & \vdots    & \vdots & \ddots & \vdots   \\
c_1       & c_2       & c_3    & \cdots & c_0      \\
\end{array}
\right].
\end{eq*}
The \emph{shift matrix} $\Shift_N$ is the circulant matrix with respect to the sequence $(c_\ell)_{\ell = 0}^{N - 1}$ with $c_1 = 1$ and $c_\ell = 0$ for all other $\ell$ reading as
\begin{eq*}
\Shift_N = 
\left[%
\begin{array}{ccccc}
  0    &   1   &   0   & \cdots & 0\\
  0    &   0   &   1   & \cdots & 0\\
\vdots &       &       & \ddots & 1\\
  1    &   0   &   0   & \cdots & 0
\end{array}%
\right].
\end{eq*}
According to \cite[eq. (3.1.4)]{bib:Davis_CirculantMatrices}, every circulant matrix can be rewritten as
\begin{eq*}
A = \sum_{\ell = 0}^{N - 1} c_\ell \Shift_N^\ell.
\end{eq*}
This class of matrices has the following properties regarding their eigenvalues and eigenvectors.
\begin{thm}[Diagonalization of Circulant Matrices, see \mbox{\cite[Theorem 3.2.2]{bib:Davis_CirculantMatrices}}
as well as \mbox{\cite[Section 3.1]{bib:Gray_ToeplitzCirculantMatrices}}]\label{thm:circulant}
Let $A \in \cL(\cH_N)$ be a circulant matrix with respect to the \dra{complex sequence $(c_k)_{k = 0}^{N - 1} $}.
Then, $A$ is diagonalizable with
\begin{eq*}
A = \QFT_N^\dagger \cdot \Lambda \cdot \QFT_N,
\end{eq*}
where $\QFT_N$ represents \dra{again} the quantum Fourier transform and $\Lambda = \diag(\lambda_0, \dots, \lambda_{N - 1})$ with
\begin{eq*}
\lambda_m = \sum_{j = 0}^{N - 1} c_j \omega^{jm} \text{~for all~} m = 0, \dots, N - 1.
\end{eq*}
\end{thm}
It should be emphasized that the original results in \cite{bib:Davis_CirculantMatrices} and \cite{bib:Gray_ToeplitzCirculantMatrices} are formulated with respect to the discrete Fourier transform.
As we are only \dra{interested} in its quantum counterpart we rewrote the statement in our sense.
\cref{thm:circulant} means that the vectors $\ket{\QFT_N m} = \frac{1}{\sqrt{N}} \sum_{j = 0}^{N - 1} \omega^{jm} \ket{j}$ form a set of eigenvectors \emph{for all} circulant matrices.
The associated eigenvalues read as the quantum Fourier transform of the vector $(c_0, \dots, c_{N - 1}) \in \bC^N$.

Interestingly, circulant matrices have been used previously in the context of quantum Fourier transform in \cite{bib:TorosovVitanov_QFTCirculant}.
Our work does however not make use of the results therein.
%
%
\section{Exponential of Projection-Based \dra{Tensor} Decompositions}\label{sec:proj}
In this section we propose a formula for the operator exponential of projection-based \dra{tensor} decompositions.
The following theorem serves as our main result in this section.
%
%
\begin{thm}\label{thm:projection_tensor_exp}
Let finite dimensional, complex Hilbert spaces $\cH$ and $\cH'$ be given.
Consider for $N \in \bN$ the operators $(P_j)_{j = 0}^{N - 1} \subseteq \cL(\cH)$ and $(\Ham_j)_{j = 0}^{N - 1} \subseteq \cL(\cH')$ that fulfill the following conditions:
\begin{enum}
\item\label{enum:sym}
The operators $(P_j)_{j = 0}^{N - 1}$ and $(\Ham_j)_{j = 0}^{N - 1}$ are symmetric.
\item\label{enum:projections}
For all $j, k = 0, \dots, N - 1$, we have $P_j P_k = \delta_{j,k} P_j$.
\item\label{enum:projection_sum}
The equation $\sum_{j = 0}^{N - 1} P_j = \id_{\cH}$ is valid.
\end{enum}
Then, the operators $(\exp(- \ic t P_j \otimes \Ham_j))_{j = 0}^{N - 1}$ commute and we have
\begin{eq*}
\dra{\exp\left( - \ic t \sum_{j = 0}^{N - 1} P_j \otimes \Ham_j \right)}
= \prod_{j = 0}^{N - 1} \exp(- \ic t P_j \otimes \Ham_j)
= \sum_{j = 0}^{N - 1} P_j \otimes \exp(- \ic t \Ham_j).
\end{eq*}
\end{thm}
\begin{proof}
First, we observe that the terms $(P_j \otimes \Ham_j)_{j = 0}^{N - 1}$ commute since
\begin{eq*}
(P_j \otimes \Ham_j)(P_k \otimes \Ham_k) = (P_j P_k) \otimes (\Ham_j \Ham_k) = 0 = (P_k \otimes \Ham_k) \otimes (P_j \otimes \Ham_j)
\end{eq*}
for $j \neq k$.
On the one hand, \cref{lem:exp_properties} yields that $\exp(- \ic t P_j \otimes \Ham_j)$ commute and 
\begin{eq*}
\exp\left(- \ic t \sum_{j = 0}^{N - 1} P_j \otimes \Ham_j\right) = \prod_{j = 0}^{N - 1} \exp(- \ic t P_j \otimes \Ham_j).
\end{eq*}
This shows the first part of the assertion.

On the other hand, we obtain inductively
\begin{eq*}
\left(\sum_{j = 0}^{N - 1} P_j \otimes \Ham_j\right)^k = \sum_{j = 0}^{N - 1} P_j \otimes \Ham_j^k
\end{eq*}
for all \dra{integers} $k \geq 1$.
Using \cref{eq:defn_exp} we get
\begin{eq*}
\exp\left(- \ic t \sum_{\dra{j = 0}}^{N - 1} P_j \otimes \Ham_j\right)
&= \id_{\cH \otimes \cH'} + \sum_{k = 1}^\infty \dra{\frac{(- \ic t)^k}{k!}} \left(\sum_{j = 0}^{N - 1} P_j \otimes \Ham_j^k \right)\\
&= \id_{\cH} \otimes \id_{\cH'} + \sum_{j = 0}^{N - 1} \sum_{k = 1}^\infty \dra{\frac{(- \ic t)^k}{k!}} P_j \otimes \Ham_j^k\\
&= \id_{\cH} \otimes \id_{\cH'} + \sum_{j = 0}^{N - 1} P_j \otimes \left( \sum_{k = 1}^\infty \frac{(- \ic t)^k}{k!} \Ham_j^k \right)\\
&= \id_{\cH} \otimes \id_{\cH'} + \sum_{j = 0}^{N - 1} \left(P_j \otimes (\exp(- \ic t \Ham_j) - \id_{\cH'}) \right)\\
&= \id_{\cH} \otimes \id_{\cH'} - \sum_{j = 0}^{N - 1} P_j \otimes \id_{\cH'} + \sum_{j = 0}^{N - 1} P_j \otimes \exp(- \ic t \Ham_j)\\
&= \sum_{j = 0}^{N - 1} P_j \otimes \exp(- \ic t \Ham_j).
\end{eq*}
In the last step, we made use of \cref{enum:projection_sum}.
This yields the remaining part of the assertion.
\end{proof}
\dra{In this article, decompositions of the form $\sum_{j = 0}^{N - 1} P_j \otimes \Ham_j$, where $P_j$, $\Ham_j$ fulfill the conditions in \cref{thm:projection_tensor_exp} are called \emph{projection-based tensor decompositions}.}
Clearly, the condition \cref{enum:projections} in \cref{thm:projection_tensor_exp} implies that the operators $(P_j)_{j = 0}^{N - 1}$ \dra{are} projections.
\dra{These decompositions} can be interpreted in two ways:
On the one hand, they are a generalization of a block diagonal matrix.
To see this, take a basis $(\ket{j})_{j = 0}^{\dim \cH - 1}$ and set $P_j = \ketbra{j}$.

On the other hand, they are essentially a generalization of the diagonalization of symmetric matrices.
To see this, let a symmetric operator $\Ham \in \cL(\cH)$ be given for some $N$-dimensional, complex Hilbert space $\cH$.
Then, there exist eigenvalues $(\lambda_j)_{j = 0}^{N - 1}$ and corresponding eigenvectors $(\ket{v_j})_{j = 0}^{N - 1}$ such that
\begin{eq*}
\Ham = \sum_{j = 0}^{N - 1} \lambda_j \ketbra{v_j}.
\end{eq*}
In this sense we can rewrite $\cH \simeq \bC \otimes \cH$ and define $P_j := \ketbra{v_j}$ and $\Ham_j := \lambda_j$ for $j = 0, \dots, N - 1$.
Then the conditions on the projections are verified and the symmetry of $\Ham_j$ is identical \dra{to the eigenvalues to be real}.

\dra{Occasionally, we make use of the following special case.}
%
%
\begin{cor}\label{cor:projection_tensor_exp_lemma}
Let finite dimensional, complex Hilbert spaces $\cH$ and $\cH'$ be given.
Take a symmetric operator $\Ham \in \cL(\cH')$ and a symmetric projection operator $P \in \cL(\cH)$.
For all $t \in \bR$ holds
\begin{eq*}
\exp\left(- \ic t P \otimes \Ham \right) = (\id_{\cH} - P) \otimes \id_{\cH'} + P \otimes \exp(- \ic t \Ham).
\end{eq*}
\end{cor}
\begin{proof}
We apply \cref{thm:projection_tensor_exp} with $N = 2$, $P_0 := P$, $P_1 = \id_{\cH} - P$ and $\Ham_0 := \Ham$, $\Ham_1 := 0$.
Clearly, $(P_j)_{j = 0, 1}$, $(\Ham_j)_{j = 0, 1}$ are symmetric operators in their respective spaces.
By definition holds $P_0 + P_1 = \id_{\cH}$, and we have
\begin{eq*}
P_0 P_1 = P (\id_{\cH} - P) = P - P^2 = 0 = P_1 P_0
\end{eq*}
as well as
\begin{eq*}
P_1^2 = (\dra{\id_{\cH}} - P)^2 = \id_{\cH} - 2P + P^2 = \id_{\cH} - P = P_1.
\end{eq*}
Hence, \dra{\cref{thm:projection_tensor_exp} yields}
\begin{eq*}
\exp(- \ic t P \otimes \Ham)
&= \exp\left(- \ic t \sum_{j = 0}^1 P_j \otimes \Ham_j\right)
= \sum_{j = 0}^1 P_j \otimes \exp(-\ic t \Ham_j)\\
&= (\id_{\cH} - P) \otimes \id_{\cH'} + P \otimes \exp(- \ic t \Ham).
\end{eq*}
This ends the proof.
%
%
\end{proof}
\dra{The previous results are illustrated with the following example.}
%
%
\begin{ex}[Controlled NOT gate]\label{ex:cnot}
Consider the controlled \NOT{} gate
\begin{eq*}
\CNOT = \left[\begin{array}{cccc} 1 & 0 & 0 & 0\\ 0 & 1 & 0 & 0\\ 0 & 0 & 0 & 1\\ 0 & 0 & 1 & 0  \end{array}\right].
\end{eq*}
Its behavior can be described as follows:
If the first qubit takes the value $\ket{1}$, then the second qubit is flipped, which is the application of the \NOT{} gate.
Otherwise, no change is performed on the second qubit.
In other words, it has the following output.
\begin{eq*}
\CNOT \ket{0}\ket{x} = \ket{0}\ket{x} \text{~and~} \CNOT \ket{1}\ket{x} = \ket{1}\ket{1-x} = \ket{1}\ket{\NOT x} \text{~for all~} x \in \{0,1\}.
\end{eq*}
It is straightforward to see
\begin{eq}\label{eq:cnot_decomp}
\CNOT = \ket{0}\bra{0} \otimes \id_2 + \ket{1}\bra{1} \otimes \NOT.
\end{eq}
%
%
We write the \NOT{} gate as the exponential of a Hamiltonian.
For this sake, we take $\ket{\pm} = \frac{1}{\sqrt{2}}(\ket{0} \pm \ket{1})$.
\dra{It is straightforward to verify}
\begin{eq*}
\NOT = \ket{+}\bra{+} - \ket{-}\bra{-}.
\end{eq*}
\dra{\cref{cor:projection_tensor_exp_lemma}} yields $\NOT = \exp(-\ic t \Ham(\NOT))$ with $t = \pi$ and $\Ham(\NOT) = \ket{-}\bra{-}$.
Hence, we rewrite the controlled \NOT{} gate as
\begin{eq*}
\CNOT = (\dra{\id_2} - \ket{1}\bra{1}) \otimes \id_2 + \ket{1}\bra{1} \otimes \exp(-\ic t \Ham(\NOT)) = \exp(-\ic t \ket{1}\bra{1} \otimes \ket{-} \bra{-}).
\end{eq*}
\end{ex}

In the upcoming result, we generalize \cref{thm:projection_tensor_exp} to incorporate transformations within the Hamiltonians.
%
%
\begin{thm}\label{thm:projection_trafo_tensor_exp}
Let finite dimensional, complex Hilbert spaces $\cH$ and $\cH'$ be given.
Consider for $N \in \bN$ the operators $(P_j)_{j = 0}^{N - 1}, S \subseteq \cL(\cH)$ and $(\Ham_j)_{j = 0}^{N - 1}, (T_j)_{j = 0}^{N - 1} \subseteq \cL(\cH')$ that fulfill the following conditions:
\begin{enum}
\item The operators $(P_j)_{j = 0}^{N - 1}$ and $(\Ham_j)_{j = 0}^{N - 1}$ are symmetric.
\item For all $j, k = 0, \dots, N - 1$, we have $P_j P_k = \delta_{j,k}P_j$.
\item The equation $\sum_{j = 0}^{N - 1} P_j = \id_{\cH}$ is valid.
\item The operators $(T_j)_{j = 0}^{N - 1}$ and $S$ are unitary.
Moreover, for all $j = 0, \dots, \dra{N - 1}$ holds $T_j = \exp(-\ic t \Ham(T_j))$.
\end{enum}
Then we have
\begin{eq*}
\exp\left(-\ic t \sum_{j = 0}^{N - 1} \left(S^\dagger P_j S\right) \otimes \left(T_j^\dagger \cdot \Ham_j \cdot T_j\right)\right)
= (S \otimes \id_{\cH'})^\dagger T^\dagger U T (S \otimes \id_{\cH'})
\end{eq*}
with
\begin{eq*}
T := \exp\left(-\ic t \sum_{j = 0}^{N - 1} P_j \otimes \Ham(T_j)\right) \text{~and~}
U := \exp\left(-\ic t \sum_{j = 0}^{N - 1} P_j \otimes \Ham_j\right).
\end{eq*}
\end{thm}
\begin{proof}
As the operators $(S^\dagger P_j S)_{j = 0}^{N - 1}$ and $(T_j^\dagger \Ham_j T_j)_{j = 0}^{N - 1}$ fulfill the criteria in \cref{thm:projection_tensor_exp}, we obtain
\begin{eq}\label{eq:tmp_proj_trafo_decomp}
\exp\left(- \ic t \sum_{j = 0}^{N - 1} (S^\dagger P_j S)\! \right. & \left.\otimes\, T_j^\dagger \Ham_j T_j \right)
= \sum_{j = 0}^{N - 1} (S^\dagger P_j S) \otimes \exp(-\ic t T_j^\dagger \Ham_j T_j)\\
&= (S \otimes \id_{\cH'})^\dagger \left(\sum_{j = 0}^{N - 1} P_j \otimes \exp(-\ic t T_j^\dagger \Ham_j T_j)\right) (S \otimes \id_{\cH'}).
\end{eq}
By \cref{lem:exp_properties} we get
\begin{eq*}
\exp(-\ic t T_j^\dagger \Ham_j T_j) = T_j^\dagger \exp(-\ic t \Ham_j) T_j.
\end{eq*}
For operators $(U_j)_{j = 0}^{N - 1},(V_j)_{j = 0}^{N - 1} \subseteq \cL(\cH)$ it is straightforward to show
\begin{eq*}
\sum_{j = 0}^{N - 1} P_j \otimes V_j U_j = \left(\sum_{j = 0}^{N - 1} P_j \otimes V_j\right)\left(\sum_{j = 0}^{N - 1} P_j \otimes U_j\right).
\end{eq*}
The application of this result on the middle factor in \cref{eq:tmp_proj_trafo_decomp} yields
\begin{eq*}
\sum_{j = 0}^{N -1} P_j &\otimes \exp(-\ic t T_j^\dagger \Ham_j T_j)
= \sum_{j = 0}^{N - 1} P_j \otimes T_j^\dagger \exp(-\ic t \Ham_j) T_j\\
&= \left( \sum_{j = 0}^{N - 1} P_j \otimes T_j^\dagger \right) \left( \sum_{j = 0}^{N - 1} P_j \otimes \exp(-\ic t \Ham_j) T_j \right)\\
&= \left( \sum_{j = 0}^{N - 1} P_j \otimes T_j \right)^\dagger \left( \sum_{j = 0}^{N - 1} P_j \otimes \exp(-\ic t \Ham_j) \right) \left( \sum_{j = 0}^{N - 1} P_j \otimes T_j \right).
\end{eq*}
In combination with $T_j = \exp(-\ic t \Ham(T_j))$, the application of \cref{thm:projection_tensor_exp} on each factor yields the assertion.
\end{proof}
The precise purpose of \cref{thm:projection_trafo_tensor_exp} will become clear later.
Nevertheless, we want to provide an application of it and return for this sake to \cref{ex:cnot}.

%
%
\begin{ex}[Alternative representation of \CNOT]\label{ex:cnot_return}
As \dra{demonstrated} in \cref{ex:cnot}, we have $\Ham(\CNOT) = \ket{1}\bra{1} \otimes \ket{-}\bra{-}$.
We apply \cref{thm:projection_trafo_tensor_exp}.
With $\ket{-} = H\ket{1}$, \dra{where $H$ is the Hadamard gate (cf. \cite[eq. (1.14)]{bib:NielsenChuang})}, we obtain for instance with \dra{$S = \id_2$} and \dra{$T_j = H$} for $j = 0, 1$ that
\begin{eq*}
\ket{1}\bra{1} \otimes \ket{-}\bra{-}
= \ket{1}\bra{1} \otimes (H\ket{1}\bra{1}H^\dagger).
\end{eq*}
Then, we \dra{set} $t = \pi$ and $\exp(-\ic t \ketbra{1}) = Z$.
\dra{Interchanging} the roles of the qubits \dra{yields}
\dra{%
\begin{eq*}
\exp(-\ic t \ket{1}\bra{1} \otimes (H\ket{1}\bra{1}H))
&= (\id_2 \otimes H) \exp(-\ic t \ket{1}\bra{1} \otimes \ket{1}\bra{1}) (\id_2 \otimes H)\\
&= (\id_2 \otimes H)(\ketbra{0} \otimes \id_2 + \ketbra{1} \otimes Z)(\id_2 \otimes H)\\
&= (\id_2 \otimes H)\CZ(\id_2 \otimes H),
\end{eq*}%
}%
where $CZ$ is the controlled $Z$ gate.

Alternatively, one could choose $S = H$ and $T_j = H$ for $j = 0, 1$.
This yields
\dra{%
\begin{eq*}
\CNOT
&= \exp\left(-\ic t (H\ketbra{-}H) \otimes (H \ketbra{1} H)\right)\\
&= (H \otimes \id_2) (\id_2 \otimes H) \exp(-\ic t \ketbra{-} \otimes \ketbra{1}) (\id_2 \otimes H) (H \otimes \id_2)\\
&= (H \otimes H) \exp(-\ic t \ketbra{-} \otimes \ketbra{1}) (H \otimes H).
\end{eq*}%
}%
In the last term, the factor in the middle is a \CNOT{} gate with the roles of control and data reversed.
These results are well-known in the quantum computing literature (cf. \cite[Exercises \dra{4.17}, 4.20]{bib:NielsenChuang}) and are subsumed in \cref{fig:cnot_equiv}.
%
%
\dra{%
\begin{figure}[h!]
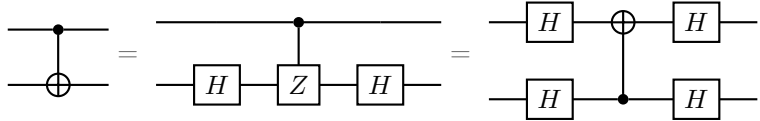

\centering
\includestandalone{\imgpath /cnot_equiv}
\caption{Alternative representations of the \CNOT{} gate.}\label{fig:cnot_equiv}
\end{figure}%
}%
\end{ex}
%
%
\section{Alternative Quantum Phase Estimation}\label{sec:alternative_qpe}
After \dra{these} preparations we now come to the main subject of our investigation---the quantum phase estimation.
Let an operator $U \in \cL(\cH)$ acting on $u$ qubits together with an eigenvector $\ket{\psi}$ and corresponding eigenvalue $\exp( \ic 2\pi \dra{\varphi})$, $\dra{\varphi} \in [0,1)$ be given.
The goal is to find (an approximation) of the phase $\varphi$.

\begin{figure}[h!]
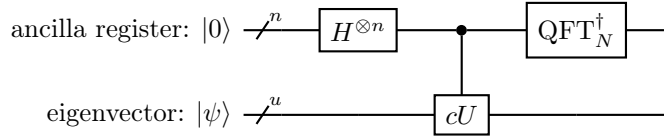

\centering
\includestandalone{\imgpath /circuit_classical_qpe}
\caption{Circuit of the quantum phase estimation.}\label{fig:classical_qpe}
\end{figure}

This can be done using \emph{quantum phase estimation}.
For this sake, an ancilla register with $n$ qubits is considered.
Here, as a reminder, we write $N = 2^n$.
Then, the circuit in \cref{fig:classical_qpe} is applied.
%
%

\dra{The} operator \dra{$cU \in \cL(\cH_N \otimes \cH)$} is called the \emph{controlled $U$ gate} and is defined by 
\begin{eq}\label{eq:defn_controlled_u}
cU\ket{j}\ket{x} := \ket{j}\ket{U^j x} \dra{\text{~for all~} j = 0, \dots, N - 1}.
\end{eq}
\dra{Let} the ancilla \dra{register be} initialized with $\ket{0}$ and \dra{assume} $N \varphi$ \dra{to be} an integer.
\dra{Then} the result after applying the quantum phase estimation in \cref{fig:classical_qpe} is $\ket{N\varphi}\ket{\psi}$.
\dra{The measurement of the ancilla register} then yields the phase as classical information.
In practice, however $N\varphi$ is not an integer and hence one obtains a superposition and a probabilistic measurement as discussed, e.g., in \cite[Section 5.2]{bib:NielsenChuang}.
For the sake of exposition we assume in the rest of this article $N \varphi$ to be an integer for our choice of $n$.
\revc{Additionally, in this article, we do not consider potential physical limitations related to the number of qubits used in the ancilla register, as discussed in \cite{bib:CaoCao_PlanckConstantQuantumFourierTransform}.
Instead, our focus lies solely on the mathematical derivations.}{}

\begin{figure}[h!]
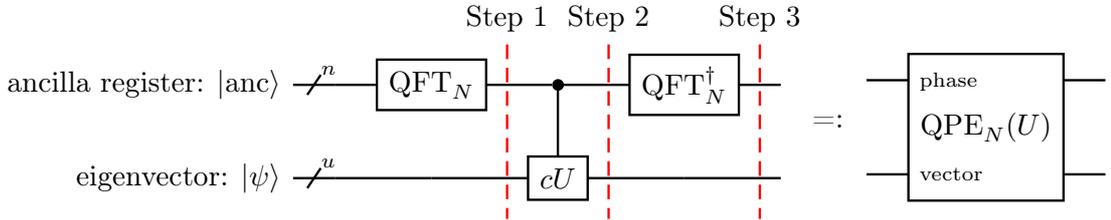

\centering
\includestandalone[width = \textwidth]{\imgpath /circuit_alternative_qpe}
\caption{Left: Circuit for the alternative quantum phase estimation. Right: Short hand representation.}\label{fig:circuit_alternative_qpe}
\end{figure}

Next, we change the quantum phase estimation \dra{depicted in \cref{fig:classical_qpe}} by substituting the application of the Hadamard transforms on every qubit to a quantum Fourier transform.
This yields the circuit depicted in \cref{fig:circuit_alternative_qpe} (left), which we will refer to as \emph{alternative quantum phase estimation}.

Occasionally, we may abbreviate this operation by the multi-qubit gate, in \cref{fig:circuit_alternative_qpe} (right), where we note which input register is devoted to the ancilla register (phase) and which one to the eigenvector \dra{(vector)}.
This is \dra{a} slight abuse of notation, as it refers to our alternative version.
Moreover, we only assume that this gate realizes the operator.
No assumption on its inner working is made, though.
%
%

Of course, for $\ket{\anc} = \ket{0}$ it is straightforward to verify $H^{\otimes n}\ket{0} = \QFT_N\ket{0}$.
In this case both circuits provide the same result.
Hence, our modification is in this sense equivalent to the traditional one.
However, the application of the quantum Fourier transform is computationally more expensive than the Hadamard transform and hence, no performance gain can be claimed.

Nevertheless, we want to analyze the action of the circuit in \cref{fig:circuit_alternative_qpe} and derive a representation of a corresponding Hamiltonian using the results in \cref{sec:proj} in the upcoming subsection.
%
%
\subsection{Action of the Alternative Quantum Phase Estimation}
Given an eigenvector $\ket{\psi}$ of $U$ with corresponding eigenvalue $e^{\ic 2\pi \varphi}$ and an ancilla qubit $\ket{\anc}$ initialized in $\ket{j}$, $j \in \{0, \dots, N - 1\}$.
\dra{We} consider the actions of the circuit in \cref{fig:circuit_alternative_qpe} after each of the steps marked therein.
For this sake let $\omega := e^{\ic \frac{2\pi}{N}}$ and let $\ket{\Step\ k}$ be the quantum state after the application of Step $k$.
\\[1em]\textbf{Step 1.}
The application of the quantum Fourier transform on the ancilla qubit yields
\begin{eq*}
\ket{\mathrm{Step}\ 1}
= (\QFT_N \otimes \id_\cH)\ket{j}\ket{\psi}
= \ket{\QFT_N j}\ket{\psi}
\dra{=} \frac{1}{\sqrt{N}} \sum_{k = 0}^{N - 1} \omega^{j k} \ket{k}\ket{\psi}.
\end{eq*}
\\[1em]\textbf{Step 2.}
Next, the controlled $U$ gate is applied, which gives
\begin{eq*}
\ket{\Step\ 2}
&= cU \ket{\Step\ 1} = \frac{1}{\sqrt{N}} \sum_{k = 0}^{N - 1} \omega^{jk} cU(\ket{k} \ket{\psi})
= \frac{1}{\sqrt{N}} \sum_{k = 0}^{N - 1} \omega^{jk} \ket{k} \ket{U^k \psi}\\
&= \frac{1}{\sqrt{N}} \sum_{k = 0}^{N - 1} \omega^{jk}  e^{\ic k 2 \pi \varphi } \ket{k}\ket{\psi}.
\end{eq*}
\\[1em]\textbf{Step 3.}
The application of the inverse quantum Fourier transform to the ancilla register yields
\begin{eq*}
\ket{\Step\ 3}
&= (\QFT_N^\dagger \otimes \id_\cH) \ket{\mathrm{Step} 2}
= \frac{1}{\sqrt{N}}\sum_{k = 0}^{N - 1} \omega^{j k} e^{\ic k 2 \pi \varphi} \QFT_N^\dagger\ket{k}\ket{\psi}\\
&= \frac{1}{N} \sum_{k = 0}^{N - 1} \sum_{m = 0}^{N - 1} \omega^{jk} \omega^{-km} e^{\ic k 2\pi \varphi} \ket{m}\ket{\psi}
= \frac{1}{N} \sum_{m = 0}^{N - 1} \left( \sum_{k = 0}^{N - 1} (\omega^j  \omega^{-m} e^{\ic 2\pi \varphi})^k \right)\ket{m}\ket{\psi}.
\end{eq*}
With $\omega = e^{\ic \frac{2 \pi}{N}}$ we get $\omega^j \omega^{-m} e^{\ic 2\pi \varphi} = e^{\ic \frac{2\pi}{N}(j - m + N\varphi)}$.
Hence, we \dra{obtain} using \cref{apx:aux}\cref{enum:aux_0} in the appendix that
\begin{eq*}
\sum_{k = 0}^{N - 1} (\omega^j \omega^{-m} e^{\ic 2\pi \varphi})^k = \left\{\begin{array}{cl}N, &\text{if~} m \equiv j + N \varphi \revb{\hspace*{-1ex}\pmod{N}} \\
0, &\text{else.}\end{array}\right.
\end{eq*}
This yields eventually 
\begin{eq*}
\ket{\mathrm{Step} 3} = \ket{j + N \varphi \revb{\hspace*{-1ex}\pmod{N}}} \ket{\psi}.
\end{eq*}
Thus, the \dra{alternative QPE} with input $\ket{j}\ket{\psi}$ yields in the ancilla register the value of \dra{$N \varphi$} shifted by $j$ \revb{in the rest class ring}.
%
%
\subsection{Hamiltonian of the Alternative Quantum Phase Estimate}
As a next step, we want to use the results in \cref{sec:proj} to represent our alternative QPE by a Hamiltonian operator.
\dra{First,} we rewrite the controlled $U$ gate using \cref{thm:projection_tensor_exp}.
By its definition in \cref{eq:defn_controlled_u} it is straightforward to verify
\begin{eq*}
cU = \sum_{j = 0}^{N - 1} \ket{j}\bra{j} \otimes U^j.
\end{eq*}
Let for the remainder of the work $\Ham(U)$ be a Hamiltonian of $U$ with $U = \exp(-\ic t \Ham(U))$.
Then we have
\begin{eq*}
U^j = \exp(-\ic t \Ham(U))^j = \exp(- \ic t j \Ham(U))
\end{eq*}
by \cref{lem:exp_properties}.
This yields with $P_j = \ketbra{j}$ and $\Ham_j = j \Ham(U)$ in \cref{thm:projection_tensor_exp} that
\begin{eq*}
cU
&= \sum_{j = 0}^{N - 1} P_j \otimes \exp(- \ic t j \Ham(U))
= \exp\left( - \ic t \sum_{j = 0}^{N - 1} P_j \otimes (j \Ham(U)) \right)\\
&= \exp\left(-\ic t \left(\sum_{j = 0}^{N - 1} j \ketbra{j}\right) \otimes \Ham(U) \right).
\end{eq*}
In other words, if we are given the Hamiltonian of an operator $U$, we can generate the Hamiltonian of the associated controlled $U$ gate by forming the tensor product with the multiplier
\begin{eq*}
C_N := \sum_{j = 0}^{N - 1} j \ketbra{j} = \operatorname{diag}(0, 1, \dots, N - 1).
\end{eq*}
Hence, we may as well write $\Ham(cU) = C_N \otimes \Ham(U)$.
Occasionally, we refer to $C_N$ as \emph{control multiplier}.

Next, we write the operator encoded in the circuit in \cref{fig:circuit_alternative_qpe} as the matrix exponential of a Hamiltonian.
Let $\QPE_N(U)$ denote the operator depicted therein.
Then we get by its definition
\begin{eq}\label{eq:alt_qpe}
\QPE_N(U) := (\QFT_N^\dagger \otimes \id_\cH) \cdot cU \cdot (\QFT_N \otimes \id_\cH).
\end{eq}

Using \cref{thm:projection_trafo_tensor_exp} we get
\begin{eq*}
\QPE_N(U)
= (\QFT_N^\dagger \otimes \dra{\id_\cH}) \cdot \exp(-\ic t C_N \otimes \Ham_j) \cdot (\QFT_N \otimes \dra{\id_\cH})
= \exp(-\ic t Q_N \otimes \dra{\id_\cH})
\end{eq*}
with
\begin{eq*}
Q_N = \QFT_N^\dagger C_N \QFT_N. 
\end{eq*}
In other words, we get $\Ham(\QPE_N(U)) = Q_N \otimes \Ham(U)$.
Analogously, we might call $Q_N$ the \emph{QPE multiplier}.
That means that the application of the alternative QPE can be represented as the tensor product of the QPE multiplier with the Hamiltonian of the target operator.
\dra{%
\begin{rem}\label{rem:qpe_inverse}
It is straightforward to see from the circuit in \cref{fig:circuit_alternative_qpe} and its formula representation in \cref{eq:alt_qpe} that $\QPE_N(U)^\dagger = \QPE_N(U^\dagger)$.
Hence, the inversion of the alternative QPE is the same as the QPE of the inverse target operator.
Also, it is straightforward to show for two commuting, unitary operators $U_0, U_1 \in \cL(\cH)$ that
\begin{eq*}
\QPE_N(U_0 U_1)
= \QPE_N(U_0) \QPE_N(U_1)
= \QPE_N(U_1) \QPE_N(U_0).
\end{eq*}
These observations do not need to be true for the textbook QPE.
\end{rem}%
}%
In the remainder of this section we would like to derive an explicit matrix representation of this operator.
This is addressed in the following subsection.
%
%
\subsection{Properties of the QPE Multiplier}
By the definition of the control multiplier we obtain
\begin{eq*}
Q_N = \dra{\QFT_N^\dagger} C_N \QFT_N = \sum_{j = 0}^{N - 1} j \ketbra{\QFT_N^\dagger j}.
\end{eq*}
By the definition of the \dra{quantum} Fourier transform we obtain
\begin{eq*}
\ket{\QFT_N^\dagger j} = \frac{1}{\sqrt{N}}\sum_{k = 0}^{N - 1} \omega^{-jk} \ket{k}.
\end{eq*}
This yields
\begin{eq*}
Q_N = \frac{1}{N}\sum_{j, k, \ell = 0}^{N - 1} j \omega^{-jk}\omega^{j\ell} \ket{k}\bra{\ell}
= \frac{1}{N} \sum_{k, \ell = 0}^{N - 1} \left( \sum_{j = 0}^{N - 1} j \omega^{(\ell - k)j} \right)\ket{k}\bra{\ell}.
\end{eq*}
Using \cref{apx:aux}\cref{enum:aux_1} in the appendix we get
\begin{eq*}
\sum_{j = 0}^{N - 1} j (\omega^{\ell - k})^j = \left\{\begin{array}{cl}
\dfrac{N (N - 1)}{2} &\text{if~} \ell = k, \\
& \\
\dfrac{N}{\omega^{\ell - k} - 1} &\text{else}.\end{array} \right.
\end{eq*}
Then, we obtain
\begin{eq}\label{eq:qpe_mult_shift}
Q_N
&= \frac{N - 1}{2} \dra{\id_N} + \sum_{\substack{k, \ell = 0,\\ k \neq \ell}}^{N - 1} \frac{1}{\omega^{\ell - k} - 1} \ket{k}\bra{\ell}
= \frac{N - 1}{2} \dra{\id_N} + \sum_{k = 0}^{N - 1}\sum_{j = 1}^{N - 1}\frac{1}{\omega^j - 1}  \ket{k}\bra{k + j \revb{\hspace*{-1ex}\pmod{N}}}\\
&= \frac{N - 1}{2} \dra{\id_N} + \sum_{j = 1}^N \frac{1}{\omega^j - 1} \Shift_N^j.
\end{eq}
Hence, the QPE multiplier is in fact \dra{the} circulant matrix with respect \dra{to} $c_0 = \frac{N -1}{2}$ and $c_j = \frac{1}{\omega^j - 1}$ \dra{that has} the eigenvalues $0, 1, \dots, N - 1.$
%
%
\section{Recursive Quantum Phase Estimate}\label{sec:recursive_qpe}
Next, we derive a recursive decomposition of the QPE multiplier.
For this sake, we decompose the $n$-qubit ancilla register into two \dra{subregisters} each having $n_0$ respectively $n_1$ qubits with $n = n_0 + n_1$.
Analogously,  \dra{set} $N_j = 2^{n_j}$ for $j = 0, 1$ with $N = N_0 \cdot N_1$.
We start by proposing the following lemma.
\begin{lem}\label{lem:circulant_decomposition}
Let a circulant matrix $A \in \cL(\cH_N)$ be given with entries 
\begin{eq*}
a_{j,k} = c_{k - j} \text{~for all~} j, k = 0, \dots, N_0 N_1 - 1
\end{eq*}
for a sequence $(c_\ell)_{\ell = 0}^{N_0 N_1 - 1} \subseteq \bC$.
Let $\ket{v_m} = \ket{\QFT_{N_0}^\dagger m}$.
Then the following equation holds
\begin{eq*}
A
= \sum_{m = 0}^{N_0 - 1} \ketbra{v_m} \otimes \left( \sum_{\ell = 0}^{N_0 - 1} \omega^{- \ell m N_1 } A^{(\ell)} \right),
\end{eq*}
where $A^{(\ell)} \in \cL(\cH_{N_1})$ with $A^{(\ell)}_{j,k} = c_{\revb{(k - j + N_1 \ell \pmod{N})}}$ for $j, k = 0, \dots, N_1 - 1$ and $\ell = 0, \dots, N_0 - 1$ \dra{and $\omega = \exp(\ic \frac{2 \pi}{N})$}.
\end{lem}
\begin{proof}
Let $A^{(\ell)}$ be defined as above.
It is straightforward to verify that \dra{the} following identity is valid
\begin{eq*}
A
= \left[
\begin{array}{ccccc}
    A^{(0)}   & A^{(1)}       & A^{(2)}  & \cdots & A^{(N_0 - 1)}\\
A^{(N_0 - 1)} & A^{(0)}       & A^{(1)}  & \cdots & A^{(N_0 - 2)}\\
A^{(N_0 - 2)} & A^{(N_0 - 1)} & A^{(0)}  & \cdots & A^{(N_0 - 3)}\\
\vdots        & \vdots        & \vdots   & \ddots & \vdots       \\
A^{(1)}       & A^{(2)}       & A^{(3)}  & \cdots & A^{(0)}      \\
\end{array}
\right].
\end{eq*}
In other words, $A$ is the Kronecker product of the matrices $A^{(\ell)}$ with powers of the $\Shift$ matrix.
Hence, we write
\begin{eq*}
A = \sum_{\ell = 0}^{N_0 - 1}  \Shift_{N_0}^\ell \otimes A^{(\ell)}.
\end{eq*}
Next, we use the diagonalization of the $\Shift_{N_0}$.
With $\omega = \exp(\ic \frac{2 \pi}{N})$ we get $\exp(\ic \frac{2\pi}{N_0}) = \omega^{N_1}$ and set
\begin{eq*}
\ket{v_m} := \ket{\QFT_{N_0}^\dagger m} = \frac{1}{\sqrt{N_0}} \sum_{j = 0}^{N_0 - 1} \omega^{- N_1 j m} \ket{j}
\end{eq*}
\dra{to} obtain
\begin{eq*}
\Shift_{N_0} = \sum_{m = 0}^{N_0 - 1} \omega^{-m N_1} \ketbra{v_m}.
\end{eq*}
Then, we further decompose
\begin{eq*}
A
= \sum_{\ell = 0}^{N_0 - 1} \left( \sum_{m = 0}^{N_0 - 1} \omega^{-\ell m N_1} \ketbra{v_m} \right) \otimes A^{(\ell)}
= \sum_{m = 0}^{N_0 - 1} \ketbra{v_m} \otimes \left( \sum_{\ell = 0}^{N_0 - 1} \omega^{-\ell m N_1} A^{\ell} \right).
\end{eq*}
This yields the assertion.
\end{proof}
Hence, every circulant matrix can be rewritten as a \dra{projection-based tensor decomposition using rank-1 projections}.
\dra{It is worth mentioning that the matrices $A^{(\ell)}$ in \cref{lem:circulant_decomposition} do not need to be circulant themselves in general.}
The factors are associated to the inverse quantum Fourier transform of the corresponding entries of $A^{(\ell)}$.
For instance we obtain for $N_0 = 2$ the decomposition
\begin{eq*}
A = \ket{+}\bra{+} \otimes (A^{(0)} + A^{(1)}) + \ket{-}\bra{-} \otimes (A^{(0)} - A^{(1)}),
\end{eq*}
where again $\ket{+} = H\ket{0}$ and $\ket{-} = H\ket{1}$.

Next, \cref{lem:circulant_decomposition} \dra{is applied} to the QPE multiplier in the upcoming theorem.
%
%
\begin{thm}\label{thm:qpe_decomposition}
Let $Q_N$ be the QPE multiplier with respect to $n = n_0 + n_1$ qubits and $N_j = 2^{n_j}$, $j = 0, 1$ with $N = N_0 \cdot N_1$.
Then, we obtain the decomposition
\begin{eq*}
Q_N = N_0 \sum_{m = 0}^{N_0 - 1} \ketbra{v_m} \otimes (D_{\omega,N_1}^{\dagger})^m Q_{N_1} D_{\omega,N_1}^m + Q_{N_0} \otimes \id_{N_1}
\end{eq*}
with $D_{\omega,N_1} := \mathrm{diag}(1, \omega, \dots, \omega^{N_1 - 1}) = \sum_{j = 0}^{N_1 - 1} \omega^j \ketbra{j}$ and $\omega = \exp(\ic \frac{2 \pi}{N})$.
\end{thm}
\begin{proof}
As we \dra{have shown} in \cref{eq:qpe_mult_shift}, \dra{the} QPE multiplier is a circulant matrix with respect to the sequence
\begin{eq*}
c_0 = \frac{N - 1}{2} \text{~and~} c_k = \frac{1}{\omega^k - 1} \text{~for~} k = 1, \dots, N - 1.
\end{eq*}
Hence, \cref{lem:circulant_decomposition} guarantees the decomposition
\begin{eq*}
Q_N = \sum_{m = 0}^{N_0 - 1} \ketbra{v_m} \otimes \left(\sum_{\ell = 0}^{N_0 - 1} \omega^{-N_1 \ell m} A^{(\ell)}\right)
\end{eq*}
with $A^{(\ell)}_{j,k} = c_{\revb{(k - j + N_1 \ell \pmod{N})}}$.
\dra{Next, we} calculate the entries of the prefactors in the above decomposition.
Let $j, k \in \{0, \dots, N - 1\}$.
We distinguish two cases.
\\[1em]\textbf{Case $j \neq k$.}
\begin{eq}\label{eq:qpe_mult_tmp_1}
\sum_{\ell = 0}^{N_0 - 1} \omega^{-N_1 \ell m} A^{(\ell)}_{j,k} = \sum_{\ell = 0}^{N_0 - 1} \frac{\omega^{-N_1 \ell m}}{\omega^{k - j + N_1\ell} - 1}.
\end{eq}
For $z \in \bC$, $|z| = 1$ with $z, z^{N_0} \neq 1$ we have by \dra{the formula for the geometric series} the identity
\begin{eq*}
\frac{z^{N_0} - 1}{z - 1} = \sum_{t = 0}^{N_0 - 1} z^t
\text{~or equivalently~}
\frac{1}{z - 1}  = \frac{1}{z^{N_0} - 1} \sum_{t = 0}^{N_0 - 1} z^t.
\end{eq*}
Hence, we obtain with $N_0 N_1 = N$, $\omega^N = 1$ that
\begin{eq}\label{eq:qpe_mult_tmp_2}
\cref{eq:qpe_mult_tmp_1}
&= \sum_{\ell = 0}^{N_0 - 1} \frac{\omega^{-N_1 \ell m}}{\omega^{N_0(k - j)}\omega^{N_0 N_1 \ell} - 1} \sum_{t = 0}^{N_0 - 1} \omega^{(k - j) t} \omega^{N_1 \ell t}\\
&= \frac{1}{\omega^{N_0(k -j)} - 1} \sum_{t = 0}^{N_0 - 1} \omega^{(k - j)t} \sum_{\ell = 0}^{N_0 - 1} \omega^{N_1 \ell (t - m)}
= \frac{N_0}{\omega^{N_0(k-j)} - 1}\omega^{(k - j)m}.
\end{eq}
\dra{Here} we used
\begin{eq}\label{eq:aux_dft}
\sum_{\ell = 0}^{N_0 - 1} \omega^{N_1 \ell (t - m)} = \left\{ \begin{array}{cl} N_0 &\text{if~} t = m,\\
0 &\text{else},
\end{array}\right.
\end{eq}
which is guaranteed \dra{by} \cref{apx:aux}\cref{enum:aux_0}.
\\[1em]\textbf{Case $j = k$.}
In this case \dra{the identity} 
\begin{eq*}
\frac{N_0}{\omega^{N_1 \ell} - 1} = \sum_{t = 0}^{N_0 - 1} t \omega^{N_1 \ell t}
\end{eq*}
derived from \cref{apx:aux}\cref{enum:aux_1} yields
\begin{eq}\label{eq:qpe_mult_tmp_3}
\sum_{\ell = 0}^{N_0 - 1} \omega^{-N_1 \ell m} A_{j,j}^{(\ell)}
&= \frac{N_0 N_1 - 1}{2} + \sum_{\ell = 1}^{N_0 - 1} \frac{\omega^{-N_1 \ell m}}{\omega^{N_1 \ell} - 1}
= \frac{N_0 N_1 - 1}{2} + \frac{1}{N_0} \sum_{\ell = 1}^{N_0 - 1} \sum_{t = 0}^{N_0 - 1} t \omega^{N_1 \ell (t - m)}\\
&= \frac{N_0 N_1 - 1}{2} + \frac{1}{N_0} \sum_{t = 0}^{N_0 - 1} t \left(-1 + \sum_{\ell = 0}^{N_0 - 1} \omega^{N_1 \ell (t - m)}\right).
\end{eq}
With \cref{eq:aux_dft} we obtain
\begin{eq*}
\cref{eq:qpe_mult_tmp_3} = \frac{N_0 N_1 - 1}{2} - \frac{1}{N_0}\sum_{t = 0}^{N_0 - 1} t + m = \frac{N_0 N_1 - 1}{2} - \frac{N_0 (N_0 - 1)}{2 N_0} + m = N_1 \frac{N_0 - 1}{2} + m.
\end{eq*}
Hence, we get with $D_{\omega,N_1}$ defined as above
\begin{eq*}
Q_N
&= \sum_{m = 0}^{N_0 - 1} \ketbra{v_m} \otimes (N_0 (D_{\omega,N_1}^\dagger)^m Q_{N_1} D_{\omega,N_1}^m + m \id_{N_1})\\
&= N_0 \sum_{m = 0}^{N_0 - 1} \ketbra{v_m} \otimes ((D_{\omega,N_1}^\dagger)^m Q_{N_1} D_{\omega,N_1}^m) + \sum_{m = 0}^{N_0 - 1} m \ketbra{v_m} \otimes \id_{N_1}\\
&= N_0 \sum_{m = 0}^{N_0 - 1} \ketbra{v_m} \otimes ((D_{\omega,N_1}^\dagger)^m Q_{N_1} D_{\omega,N_1}^m) +  Q_{N_0} \otimes \id_{N_1},
\end{eq*}
which completes the proof.
\end{proof}
Next, we return to the quantum phase estimation of a unitary operator $U$ with associated Hamiltonian $\Ham(U)$ and apply the decomposition in \cref{thm:qpe_decomposition} to rewrite the operator.
%
%
\subsection{Composed Quantum Phase Estimation}
\begin{figure}[h!]
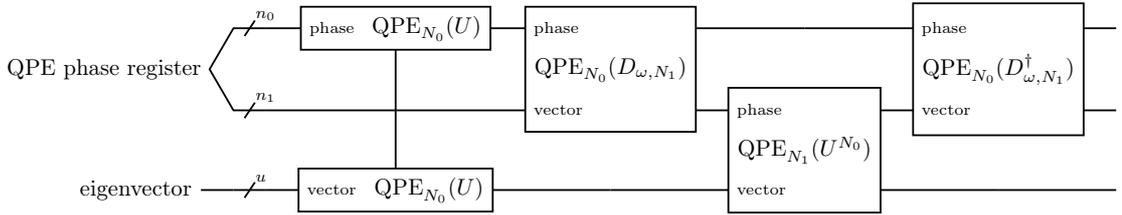

\includestandalone[width = \textwidth]{\imgpath /circuit_recursive_qpe}
\caption{Depiction of \dra{the} decomposition in \cref{eq:qpe_composition}.}\label{fig:recursive_qpe}
\end{figure}

As an attempt to decompose the quantum phase estimate into smaller operations, we use the result in \cref{thm:qpe_decomposition}.
First and foremost, we see that the operators $Q_{N_0} \otimes \id_{N_1}$ and $\ketbra{v_m} \otimes (D_{\omega,N_1}^{m \dagger} Q_{N_1} D_{\omega,N_1}^m)$ for $m = 0, \dots, N_0 - 1$ commute.
Hence, we get by subsequent use of \cref{thm:projection_trafo_tensor_exp} the following product
\begin{eq*}
\QPE_N(U)
= \exp(-\ic t Q_N \otimes U)
=  (T \otimes \id_\cH)^\dagger V (T \otimes \id_\cH) \exp(-\ic t Q_{N_0} \otimes \id_{N_1} \otimes \Ham(U))
\end{eq*}
with
\begin{eq*}
T
&= \exp\left(-\ic t \sum_{m = 0}^{N_0 - 1} \ketbra{v_m} \otimes \Ham(D_{\omega,N_1}^m) \right)
= \exp\left(-\ic t \sum_{m = 0}^{N_0 - 1} m \ketbra{v_m} \otimes \Ham(D_{\omega,N_1}) \right)\\
&= \exp\left(-\ic t Q_{N_0} \otimes \Ham(D_{\omega,N_1})\right) = \QPE_{N_0}(D_{\omega,N_1})
\end{eq*}
and
\begin{eq*}
V
&= \exp\left(- \ic t N_0 \sum_{m = 0}^{N_0 - 1} \ketbra{v_m} \otimes Q_{N_1} \otimes \Ham(U) \right)
= \exp(-\ic t\, \id_{N_0} \otimes Q_{N_1} \otimes \Ham(U^{N_0}))\\
&= \id_{N_0} \otimes \QPE_{N_1}(U^{N_0}).
\end{eq*}
In combination, this yields with $\QPE_{N_0}(D_{\omega,N_1}^\dagger) = \QPE_{N_0}(D_{\omega,N_1})^\dagger$ \dra{(see \cref{rem:qpe_inverse})} the following decomposition
\begin{eq}\label{eq:qpe_composition}
\QPE_N(U)
&= (\QPE_{N_0}(D_{\omega,N_1}^\dagger) \otimes \id_\cH) (\id_{N_0} \otimes \QPE_{N_1}(U^{N_0})) (\QPE_{N_0}(D_{\omega,N_1}) \otimes \id_\cH)\\
&\cdot \QPE_{N_0}(\id_{N_1} \otimes U ).
\end{eq}
Of course, the last term performs a QPE of $U$ with respect to $n_0$ qubits and ignores the other $n_1$ qubits in the quantum register associated to $\QPE_N(U)$.
The representation in \cref{eq:qpe_composition} is as well depicted in
\cref{fig:recursive_qpe}.
%
%

\begin{figure}[h!]
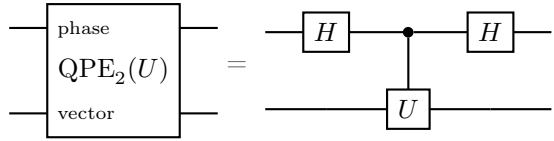

\centering
\includestandalone{\imgpath /circuit_one_qubit_qpe}
\caption{Quantum phase estimation with respect to a single phase qubit.}\label{fig:qpe_one_qubit}
\end{figure}

Let us consider for the sake of explanation the case $n_0 = 1$.
Then, we obtain for a one qubit QPE the circuit in \cref{fig:qpe_one_qubit}.
With $\QFT_2 = H$ we deduce the following equivalence in \cref{fig:qpe_one_qubit}.
%
%

\begin{figure}
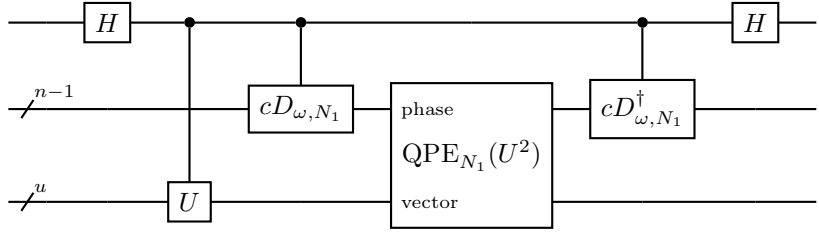

\centering
\includestandalone{\imgpath /circuit_recursive_qpe_one_qubit}
\caption{\dra{Simplification of the circuit in \cref{fig:recursive_qpe} for the} special case $n_0 = 1$.}\label{fig:recursive_qpe_one_qubit}
\end{figure}

If we chose to plug in the representation therein into the circuit in \cref{fig:recursive_qpe}, we see that due to the iterated quantum phase transform many Hadamard transforms do cancel out leading to the result in
\cref{fig:recursive_qpe_one_qubit}.
%
%
\dra{%
Of course, one might be tempted to iterate this procedure.
It is straightforward to show that the resulting circuit contains the circuits used to implement the quantum Fourier transform (cf. \cite[Fig. 5.1]{bib:NielsenChuang}).
As the demonstration is space-consuming, it is left to the curious reader.%
}%

\begin{rem}
\dra{%
For a recursive construction of a circuit, the case $n_0 = n_1 = \frac{n}{2}$ is of particular interest.
\cref{fig:recursive_qpe} contains four smaller instances of alternative quantum phase estimates.
Let us assume in this remark that the alternative QPE for $U$ and $U^{N_0}$ cost the same for the same number of ancilla qubits.
Then the overall complexity of the circuit depends on the question, whether a simplified circuit for the alternative QPE of $D_{\omega, N_1}$ can be found.
If for instance, one could find for the latter a circuit with complexity $\cO(n)$, then the algorithmic master theorem would yield an overall complexity of $\cO(n \log(n))$.
Whether such a circuit exists is an open question and hence no performance \reva{gain} is claimed.}
\end{rem}

%
%
\subsection{Nested Quantum Phase Estimation}
\begin{figure}
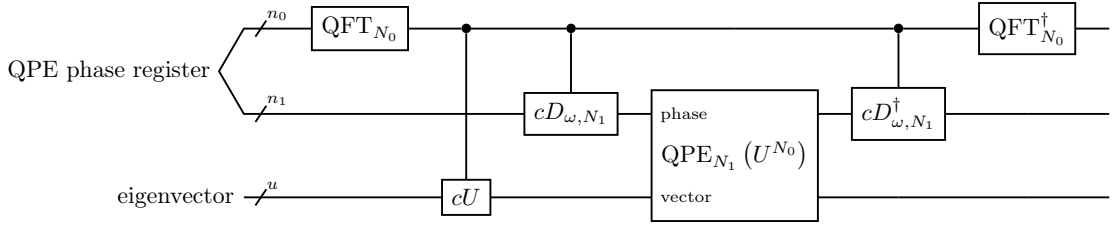

\centering
\includestandalone[width = \textwidth]{\imgpath /circuit_nested_qpe}
\caption{Nested representation.}\label{fig:qpe_nested}
\end{figure}

The recursive application of the results in the previous section might be inefficient due to the repeated application of Hadamard transforms that cancel out.
This requires in practice additional steps to simplify the resulting circuit \dra{prior to its implementation}.
However, in this subsection we want to provide an alternative decomposition of the quantum phase estimation.
By definition holds
\begin{eq*}
\ketbra{v_m} = \QFT_{N_0}^\dagger \ketbra{m} \QFT_{N_0}
\text{~and~}
Q_{N_0} = \QFT_{N_0}^\dagger C_{N_0} \QFT_{N_0}.
\end{eq*}
Hence, we can apply \cref{thm:projection_trafo_tensor_exp} and obtain
\begin{eq*}
&\exp(- \ic t Q_N \otimes \Ham(U))\\
&= \exp\left(-\ic t N_0 \sum_{m = 0}^{N_0 - 1} \ketbra{v_m} \otimes D_{\omega,N_1}^{m\dagger} Q_{N_1} D_{\omega,N_1}^m \otimes \Ham(U)
- \ic t Q_{N_0} \otimes \id_{N_1} \otimes \Ham(U) \right)\\
&= (\QFT_{N_0}^\dagger \otimes \id_{N_1} \otimes \id_\cH) \cdot W \cdot (\QFT_{N_0} \otimes \id_{N_1} \otimes \id_\cH)
\end{eq*}
with
\begin{eq*}
W = \exp\left(-\ic t N_0 \sum_{m = 0}^{N_0 - 1} \ketbra{m} \otimes D_{\omega,N_1}^{m\dagger} Q_{N_1} D_{\omega,N_1}^m \otimes \Ham(U) \reva{-\ic t} C_{N_0} \otimes \id_{N_1} \otimes \Ham(U)\right).
\end{eq*}
As the operator $C_{N_0} \otimes \dra{\id_{N_1}} \otimes \Ham(U)$ commutes with all $\ketbra{m} \otimes D_{\omega,N_1}^{m\dagger} Q_{N_1} D_{\omega,N_1}^m \reva{\otimes \Ham(U)}$ for $m = 0, \dots, N_0 -1$ we \dra{obtain} with \cref{thm:projection_trafo_tensor_exp} the product
\begin{eq*}
W
= \left(V^\dagger \exp\left(-\ic t N_0 \sum_{m = 0}^{N_0 - 1} \ketbra{m} \otimes Q_{N_1} \otimes \Ham(U)\right) V \right)\exp(-\ic t C_{N_0} \otimes \id_{N_1} \otimes \Ham(U))
\end{eq*}
with
\begin{eq*}
V = \sum_{m = 0}^{N_0 - 1} \ketbra{m} \otimes D_{\omega,N_1}^m \otimes \id_\cH = cD_{\omega,N_1} \otimes \id_\cH.
\end{eq*}
Further, we deduce
\begin{eq*}
N_0 \sum_{m = 0}^{N_0 - 1} \ketbra{m} \otimes Q_{N_1} \otimes \Ham(U)
= \id_{N_0} \otimes Q_{N_1} \otimes \Ham(U^{N_0})
\end{eq*}
and hence with the definition of the alternative quantum phase estimate that
\begin{eq*}
W = (cD_{\omega,N_1}^\dagger \otimes \id_\cH) \cdot (\id_{N_0} \otimes \QPE_{N_1}(U)) \cdot (cD_{\omega,N_1} \otimes \id_\cH) \cdot c(\id_{N_1} \otimes U).
\end{eq*}
Hereby, we note that
\begin{eq*}
(cD_{\omega,N_1} \otimes \id_\cH) \cdot c(\id_{N_1} \otimes U)
&= \left( \sum_{m = 0}^{N_0 - 1} \ketbra{m} \otimes D_{\omega,N_1}^m \otimes \id_\cH \right)
\left( \sum_{m = 0}^{N_0 - 1} \ketbra{m} \otimes \id_{N_1} \otimes U^m \right)\\
&= \sum_{m = 0}^{N_0 - 1} \ketbra{m} \otimes D_{\omega,N_1}^m \otimes U^m
= c(D_{\omega,N_1} \otimes U).
\end{eq*}
In other words $U$ and $D_{\omega,N_1}$ are simultaneously controlled by the same subregister.
This motivates the representation as a nested quantum phase estimate given in \cref{fig:qpe_nested}.
In fact the quantum phase inside with respect to $N_1$ is not controlled by the other subregister.
\dra{A closer inspection of the circuit in \cref{fig:qpe_nested} reveals that the first and last operations are the QFT and its inverse with respect to $n_0$ qubits.
All the operations in between are either independent of or controlled by these qubits.
As this resembles the circuit in \cref{fig:circuit_alternative_qpe} we call this the \emph{nested} representation.

\reva{Let us point out that both, composed and nested QPE are recursive decompositions of the alternative quantum phase estimation introduced in Section 4.}
}
%
%
%
%
\section{Conclusion and Outlook}
In this work, we proposed an alternative quantum phase estimation.
\dra{We were capable to demonstrate that this modification has indeed many interesting and convenient mathematical properties.}
\dra{For the analysis, we used formulas for the operator exponential of projection-based tensor decompositions.}

Addressing the latter, it is not clear whether such a decomposition exists for all tensor decompositions of the underlying space.
Clearly, the diagonalization is such a case, but are \dra{they} more generally available?
Even if not, we have seen that there are cases that comfortably fit into this framework.
\dra{In particular, the formulas in \cref{thm:projection_tensor_exp}, \cref{cor:projection_tensor_exp_lemma} and \cref{thm:projection_trafo_tensor_exp} seem to be} attractive for automated  compilation schemes, as it decomposes a possibly large operator into smaller ones.
This might enable a recursive build of the circuit and has in principle been demonstrated in \cref{sec:recursive_qpe}, besides no performance gain has been \dra{proven}.

The QPE is \dra{built} upon the quantum Fourier transform.
There are generalizations of the Fourier transform to \dra{nonabelian} finite groups, see \cite[Chapter 15]{bib:Terras_FourierAnalysisFiniteGroupsApplications}, where the classical Fourier transform can be interpreted as \dra{a} special case on a cyclic group.
The matrix representation of the latter using permutation groups is exactly the set of powers of the shift matrix.
As we have seen, \dra{the QPE multiplier is a} symmetric matrix that is a complex linear combination of powers of the shift matrix with eigenvalues $0, 1, \dots, N-1$.
One might be tempted to use this observation as a starting point to construct generalized quantum phase estimations upon nonabelian groups by substituting the shift with the corresponding matrix representations and adapting the coefficients.
\dra{Therein, the ancilla qubit would encode an element from the underlying group.}
\dra{All of this is however left for future investigations.}
%
%
\dra{%
\section*{Acknowledgments}
I would like to thank S\'ebastien Designolle, Patrick Gel\ss{} and Zarin Shakibaei for reading an earlier version of the manuscript and providing constructive comments.
}%

%
%
\dra{%
\section*{Funding and Competing Interests}
This study is funded by the \emph{Einstein Research Unit Perspectives of a quantum digital transformation: Near-term quantum computational devices and quantum processors}.
The author has no competing interests to declare that are relevant to the content of this article.%
}%
\section*{Data Availability Statement}
Data sharing is not applicable to this article as no datasets were generated or analyzed during the current study.

%
%
\printbibliography

@article{bib:CaoCao_PlanckConstantQuantumFourierTransform,
  title={{T}he {P}lanck {C}onstant and {Q}uantum {F}ourier {T}ransformation},
  author={Cao, Zhengjun and Cao, Zhenfu},
  journal={Cryptology ePrint Archive},
  year={2023}
}

@article{bib:HHL,
  title={Quantum algorithm for linear systems of equations},
  author={Harrow, Aram W and Hassidim, Avinatan and Lloyd, Seth},
  journal={Physical review letters},
  volume={103},
  number={15},
  pages={150502},
  year={2009},
  publisher={APS}
}

@book{bib:Terras_FourierAnalysisFiniteGroupsApplications,
  title={{Fourier Analysis on Finite Groups and Applications}},
  author={Terras, A.},
  isbn={9780521457187},
  lccn={98036455},
  series={London Mathematical Society Student Texts},
  year={1999},
  publisher={Cambridge University Press}
}

@book{bib:Hall_LieGroupsLieAlgebras,
  title={{Lie Groups, Lie Algebras, and Representations: An Elementary Introduction}},
  author={Hall, B.},
  isbn={9783319134673},
  series={Graduate Texts in Mathematics},
  year={2015},
  publisher={Springer International Publishing}
}

@article{bib:Feynman,
  title={Simulating Physics with Computers},
  author={Feynman, R. P.},
  journal={International Journal of Theoretical Physics},
  volume={21},
  number={6/7},
  year={1982}
}

@inproceedings{bib:Shor,
  title={Algorithms for quantum computation: discrete logarithms and factoring},
  author={Shor, P. W.},
  booktitle={Proceedings 35th annual symposium on foundations of computer science},
  pages={124--134},
  year={1994},
  organization={Ieee}
}

@inproceedings{bib:Grover,
  title={A fast quantum mechanical algorithm for database search},
  author={Grover, L. K.},
  booktitle={Proceedings of the twenty-eighth annual ACM symposium on Theory of computing},
  pages={212--219},
  year={1996}
}

@article{bib:Benioff_MiscroscopicQuantumMechanical,
  title={The computer as a physical system: A microscopic quantum mechanical Hamiltonian model of computers as represented by Turing machines},
  author={Benioff, P.},
  journal={Journal of statistical physics},
  volume={22},
  number={5},
  pages={563--591},
  year={1980},
  publisher={Springer}
}

@article{bib:Peres_ReversibleLogicQuantumComputer,
  title={{Reversible logic and quantum computers}},
  author={Peres, A.},
  journal={Physical review A},
  volume={32},
  number={6},
  pages={3266},
  year={1985},
  publisher={APS}
}

@article{bib:Kitaev_QPE,
  title={{Quantum measurements and the Abelian stabilizer problem}},
  author={Kitaev, A. Y.},
  journal={arXiv preprint quant-ph/9511026},
  year={1995}
}

@book{bib:NielsenChuang,
  title={{Quantum Computation and Quantum Information: 10th Anniversary Edition}},
  author={Nielsen, M.A. and Chuang, I.L.},
  isbn={9781107002173},
  lccn={2011290698},
  year={2010},
  publisher={Cambridge University Press}
}

@book{bib:Manin_ComputableNoncomputable,
  title={Computable and Noncomputable (original title in Russian: \foreignlanguage{russian}{Вычислимое и невычислимое}},
  author={Manin, Y.},
  lccn={83464240},
  year={1980},
  publisher={Sovetskoe Radio \foreignlanguage{russian}{(Сов. радио)}}
}

@article{bib:TorosovVitanov_QFTCirculant,
  title={{Design of quantum Fourier transforms and quantum algorithms by using circulant Hamiltonians}},
  author={Torosov, B. T. and Vitanov, N. V.},
  journal={Physical Review A},
  volume={80},
  number={2},
  pages={022329},
  year={2009},
  publisher={APS}
}

@book{bib:Gray_ToeplitzCirculantMatrices,
  title={Toeplitz and Circulant Matrices: A Review},
  author={Gray, R.M.},
  isbn={9781933019239},
  lccn={2006045381},
  series={Foundations and Trends in Technology},
  year={2006},
  publisher={Now Publishers}
}

@book{bib:Davis_CirculantMatrices,
  title={{Circulant Matrices}},
  author={Davis, P. J.},
  isbn={9780471057710},
  lccn={79010551},
  series={Monographs and textbooks in pure and applied mathematics},
  year={1979},
  publisher={Wiley}
}

@book{bib:Hackbusch_TensorSpacesNumericalTensorCalculus,
  title={{Tensor Spaces and Numerical Tensor Calculus}},
  author={Hackbusch, W.},
  isbn={9783030355548},
  series={Springer Series in Computational Mathematics},
  year={2019},
  publisher={Springer International Publishing}
}

%
%
\appendix
\section{Appendix}
%
%
\begin{thm}\label{apx:aux}
Let $z \in \bC$, $|z| = 1$ and $z^N = 1$ be given.
Then the following identities are valid.
\begin{enum}
\item\label{enum:aux_0}
\begin{eq}\label{eq:aux_0}
\sum_{j = 0}^{N - 1} z^j = \left\{\begin{array}{cl} N &\text{if~} z = 1,\\
0 & \text{else}. \end{array}\right.
\end{eq}
\item\label{enum:aux_1}
\begin{eq}\label{eq:aux_1}
\sum_{j = 0}^{N - 1} j z^j = \left\{\begin{array}{cl} \dfrac{N(N - 1)}{2} &\text{if~} z = 1,\\
 & \\
\dfrac{N}{z - 1} &\text{else}.\end{array}\right.
\end{eq}
\end{enum}
\end{thm}
\begin{proof}
\textbf{ad} \cref{enum:aux_0}.
If $z \neq 1$, then we get by the geometric series
\begin{eq*}
\sum_{j = 0}^{N - 1} z^j = \frac{z^N - 1}{z - 1} = 0
\end{eq*}
as $z^N = 1$.
If $z = 1$, then trivially
\begin{eq*}
\sum_{j = 0}^{N - 1} z^j = \sum_{j = 0}^{N - 1} 1 = N.
\end{eq*}
\\ \textbf{ad} \cref{enum:aux_1}.
Let again $z \neq 1$.
We use \cref{enum:aux_0}.
Then we get by differentiation and multiplication with $z$ the identity
\begin{eq}
\sum_{j = 0}^{N - 1} j z^j
&= z \frac{d}{d\xi}\left(\sum_{j = 0}^{N - 1} \xi^j\right)(z)
= z \frac{d}{d\xi}\left(\frac{\xi^N - 1}{\xi - 1} \right)(z)\\
&= z \left( N z^{N - 1} \frac{1}{z - 1} - (z^N - 1) \frac{1}{(z - 1)^2} \right)\\
&= \frac{1}{(z - 1)^2} \left( N z^{N + 1} - N z^N - z^{N+1} + z \right)
= \frac{N}{(z - 1)}.
\end{eq}
For $z = 1$ we get
\begin{eq*}
\sum_{j = 0}^{N - 1} j z^j = \sum_{j = 0}^{N - 1} = \frac{N (N - 1)}{2}.
\end{eq*}
\end{proof}
\end{document}